\documentclass[iop]{emulateapj}

\usepackage{verbatim}

\newcommand{\zwindow}{$0.5<z<1$}
\newcommand{\mcut}{10.3}

\newcommand{\msun}{$M_{\odot}$}
\newcommand{\masslimit}{$M>10^{\mcut}~M_{\odot}$}
\newcommand{\mgroup}{$M_{\rm stars}$}
\newcommand{\mgroupmcut}{$M_{\rm stars}(>10^{\mcut}~M_{\odot})$}
\newcommand{\mh}{$M_{200}$}

\begin{document}

\slugcomment{Accepted to ApJL}
\shorttitle{Stellar Mass -- Halo Mass Relation for Groups at $0.5<z<1$}
\shortauthors{Patel et al.}

\title{The Stellar Mass -- Halo Mass Relation for Low Mass X-ray Groups at $0.5<\lowercase{z}<1$ in the CDFS with CSI\altaffilmark{*}}

\author{Shannon G. Patel$^1$, Daniel D. Kelson, Rik J. Williams, John S. Mulchaey, Alan Dressler, Patrick J. McCarthy, Stephen A. Shectman}

\affil{Carnegie Observatories, 813 Santa Barbara Street, Pasadena, CA 91101, USA}
\email{patel@obs.carnegiescience.edu}

\altaffiltext{*}{This Letter includes data gathered with the 6.5 m Magellan Telescopes located at Las Campanas Observatory, Chile.}

\begin{abstract}
Since $z\sim 1$, the stellar mass density locked in low mass groups and clusters has grown by a factor of $\sim 8$.  Here we make the first statistical measurements of the stellar mass content of low mass X-ray groups at $0.5<z<1$, enabling the calibration of stellar-to-halo mass scales for wide-field optical and infrared surveys.  Groups are selected from combined $Chandra$ and $XMM-Newton$ X-ray observations in the Chandra Deep Field South (CDFS).  These ultra-deep observations allow us to identify bona fide low mass groups at high redshift and enable measurements of their total halo masses.  We compute aggregate stellar masses for these halos using galaxies from the Carnegie-Spitzer-IMACS (CSI) spectroscopic redshift survey.  Stars comprise $\sim 3-4\%$ of the total mass of group halos with masses $10^{12.8}<M_{200}/M_{\odot}<10^{13.5}$ (about the mass of Fornax and 1/50th the mass of Virgo).  Complementing our sample with higher mass halos at these redshifts, we find that the stellar-to-halo mass ratio decreases toward higher halo masses, consistent with other work in the local and high redshift universe.  The observed scatter about the stellar-halo mass relation is $\sigma \sim 0.25$~dex, which is relatively small and suggests that total group stellar mass can serve as a rough proxy for halo mass.  We find no evidence for any significant evolution in the stellar-halo mass relation since $z\lesssim 1$.  Quantifying the stellar content in groups since this epoch is critical given that hierarchical assembly leads to such halos growing in number density and hosting increasing shares of quiescent galaxies.
\end{abstract}

\section{Introduction}

The buildup of stellar mass in dark matter halos has been measured in the local universe with a variety of methods, indirectly with abundance matching techniques \citep{behroozi2013d,moster2013} and more directly with measurements of stellar and total masses of individual halos \citep{kravtsov2014}.  These stellar mass -- halo mass relations provide strong constraints on models of galaxy formation \citep[e.g.,][]{genel2014}.

While much of the attention has been focussed on the stellar content of $L^{\star}$ halos, recent work has highlighted the relevance of the higher mass scales represented by groups and clusters, where various feedback processes are predicted to operate.  \citet{kravtsov2014} for instance show discrepancies between stellar-to-halo mass ratios computed for clusters from abundance matching and from direct observations.  Meanwhile, at $z \lesssim 1$, \citet{williams2012b} measure a dramatic increase in the number density of galaxy groups over an epoch when star formation activity is greatly diminished, both in relative intensity \citep[i.e., SSFR,][]{fumagalli2012} and in the number of galaxies that display active levels of star formation at a fixed mass.  Given that these massive halos at high redshift represent sites where galaxies are quenched at levels above that in the field \citep{patel2009,patel2011}, an analysis of environmental trends provides insight into the mechanisms that contribute toward the growth of the red sequence.  Directly measuring the mass scales of these halos in which galaxies undergo quenching will provide stronger constraints on the relevant mechanisms than indirect methods for inferring halo membership and halo mass scales.

The two relevant measurements for directly probing group mass scales are the halo mass and the total group stellar mass.  At high redshift, reliable detection of galaxy groups as well as halo mass estimates require deep X-ray imaging.  We therefore utilize the ultra deep Chandra+XMM Newton group catalog in the CDFS of \citet{finoguenov2014}.  To find galaxies within these groups, we require a survey with sufficient redshift precision, which is uniform, and which has well-characterized completeness corrections to relatively low stellar masses.  The Carnegie-Spitzer-IMACS (CSI) survey \citep{kelson2014} meets these criteria and is used in this work.

We assume a cosmology with $H_0=70$~km~s$^{-1}$~Mpc$^{-1}$, $\Omega_M=0.30$, and $\Omega_{\Lambda}=0.70$.  Stellar masses are based on a \citet{kroupa2001b} IMF.

\section{Data and Analysis}

\begin{figure*}
\epsscale{1.15}  
\plottwo{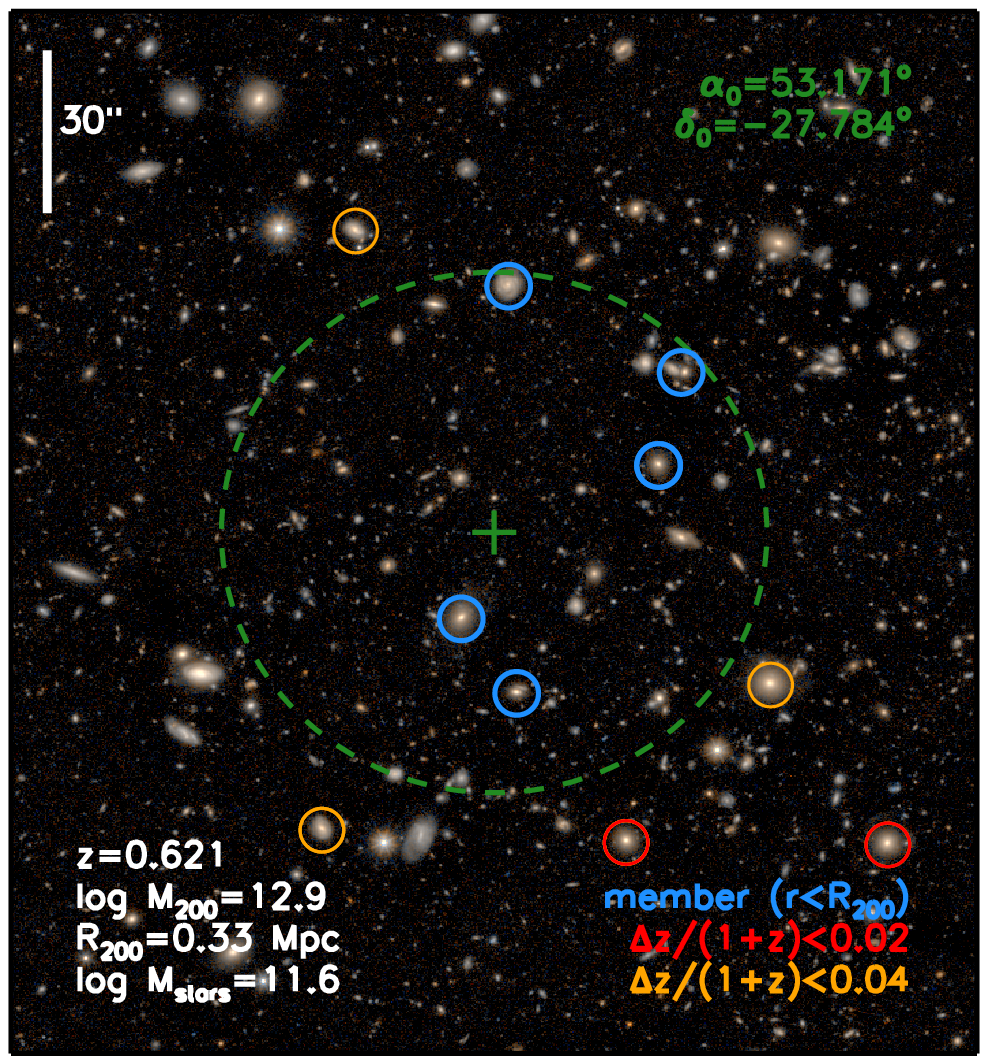}{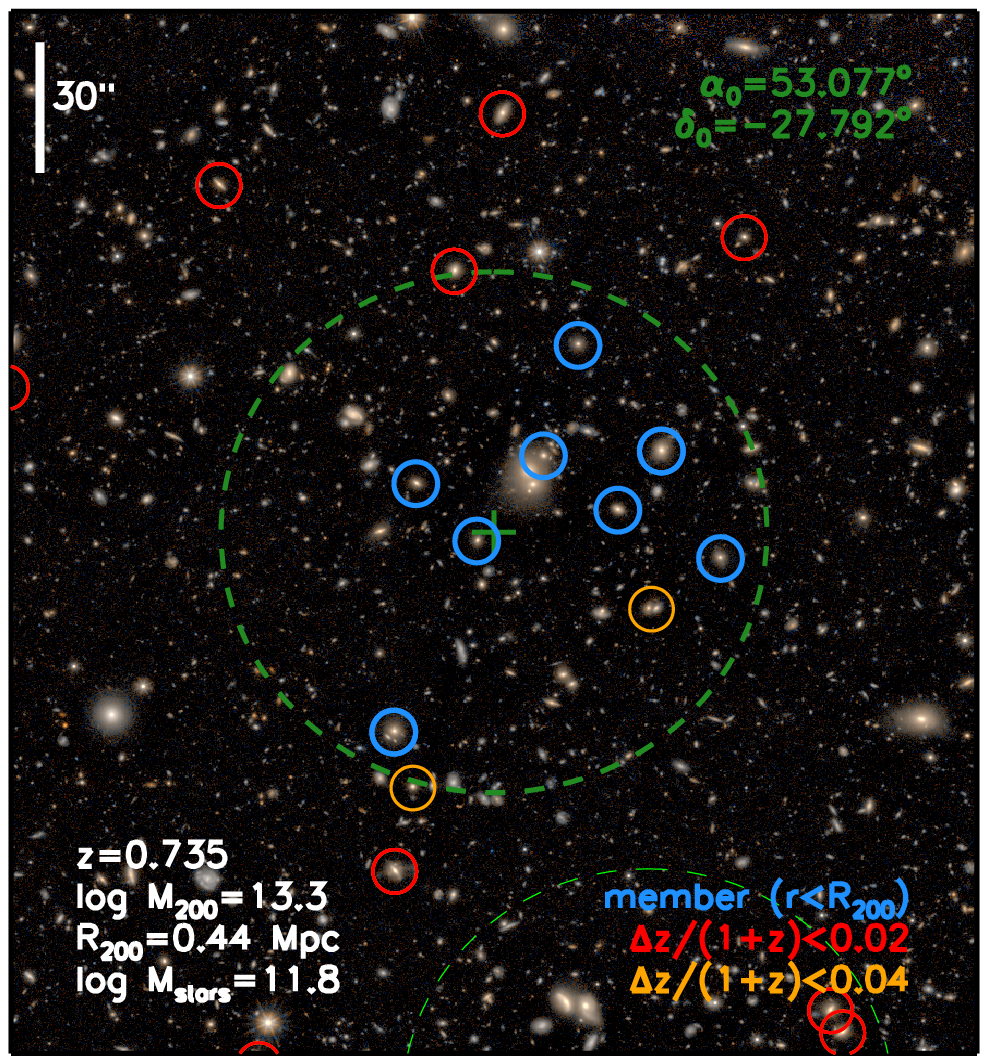}
\caption{Example X-ray detected groups in the CDFS.  The color images are constructed from HST/WFC3 $J_{125}$ and $H_{160}$ bands.  The centers of the groups are marked with a green plus sign and the virial radii, $R_{200}$, are indicated by the large dashed green circles.  Galaxies that are characterized as members of the group are circled in blue, while nearby field galaxies are circled in red and orange based on their redshift offset from the group.  Clear overdensities exist in the vicinity of the identified group centers.} \label{fig_map}
\end{figure*}

\subsection{Chandra and XMM-Newton X-ray Imaging}

We utilize combined Chandra and XMM-Newton X-ray observations in the CDFS \citep{finoguenov2014} to select groups at high redshift.  These observations represent the deepest X-ray data to date.  For the analysis presented here, we cull the list of $46$ X-ray groups from \citet{finoguenov2014} to those with redshifts in the range \zwindow\ and flags of $1$ or $2$, indicating well defined centroids and redshift determinations from followup spectroscopy.  In addition, we include only those groups with CSI coverage in the ECDFS, out of $4.8$~sq~degs in the SWIRE CDFS field.  Only $23$ groups remain in the catalog.  Total masses, $M_{200}$, given in \citet{finoguenov2014} are estimated from the $L_X-M$ relation in \citet{leauthaud2010}, which was calibrated with weak lensing based mass estimates in the COSMOS field.  Though the lowest halo masses were derived by extrapolating the latter relation, \citet{finoguenov2014} confirmed the reliability of the halo mass estimates from stacked weak lensing and clustering analyses.  $R_{200}$ values follow from these mass estimates. 

\subsection{CSI Spectroscopy and Completeness}

We utilize a redshift catalog in the CDFS constructed from new CSI observations.  The handling of the data is similar to that in the XMM-LSS field \citep{kelson2014}, though our deep DECam photometry provides for more uniform completeness as a function of magnitude.  Briefly, CSI targets were selected in the {\em Spitzer} IRAC 3.6~\micron\ band with magnitudes brighter than $21$~AB~mag and observed with the Uniform Dispersion Prism (UDP) on IMACS \citep{dressler2011} at Magellan.  Redshifts were measured by fitting the available spectrophotometry (spectrum~$+~ugrizJHK$) with SEDs generated from stellar population synthesis models.  In comparisons with high-resolution redshifts from \citet{cooper2012b} we find a biweight scatter of $\sigma_z/(1+z) \sim 0.008$ for both red and blue galaxies.  At $z=1$, the stellar mass completeness limit is $M \sim 10^{\mcut}~M_{\odot}$.

We note that group numbers $79$ and $68$ do not contain CSI galaxies above the mass cut of \masslimit\ (some galaxies may have been missed due to slit collisions and other sources of incompleteness), though both contain galaxies below this limit.  These are both the lowest mass groups in the sample with halo mass estimates of $M_h \sim 10^{12.8}$~\msun.  Also, number 63 had a large uncertainty on the stellar mass estimate (see below).  We exclude these groups from our analysis leaving a total sample of 20 X-ray groups in the CDFS.

The spectroscopic completeness for CSI in the CDFS is $\sim 50-60\%$ over $17<i<25.5$~AB~mag and a broad range of color.  To accurately account for galaxies that were not observed with the UDP or which did not have well constrained redshifts, galaxies are assigned weights according to their inverse completeness in bins of observed magnitude ($[3.6]$), observed color ($i-[3.6]$), and local source density \citep[see, e.g.,][]{kelson2014}.

Even though the CDFS has been extensively studied by other surveys, with numerous high-resolution redshifts available in the literature \citep[e.g.,][]{cooper2012b}, we refrain from combining the multiple available datasets.  The different selection functions in other work would make it difficult to produce a homogeneous dataset from the aggregate of all the available surveys, despite potentially boosting the level of completeness.  The depth and homogeneity of the CSI data simplify the analysis, and the relatively uniform $\sim 50-60\%$ completeness is high enough to robustly characterize the stellar mass content of these X-ray groups.  Lastly, while other works have used photometric catalogs to assemble or calibrate galaxy samples in groups in the CDFS \citep[e.g.,][]{giodini2009,erfanianfar2014}, our sample is exclusively spectroscopic.

\subsection{Group Membership and Aggregate Stellar Mass Calculation and Uncertainty} \label{sec_member}

\begin{figure*}
\epsscale{1.2}
\plotone{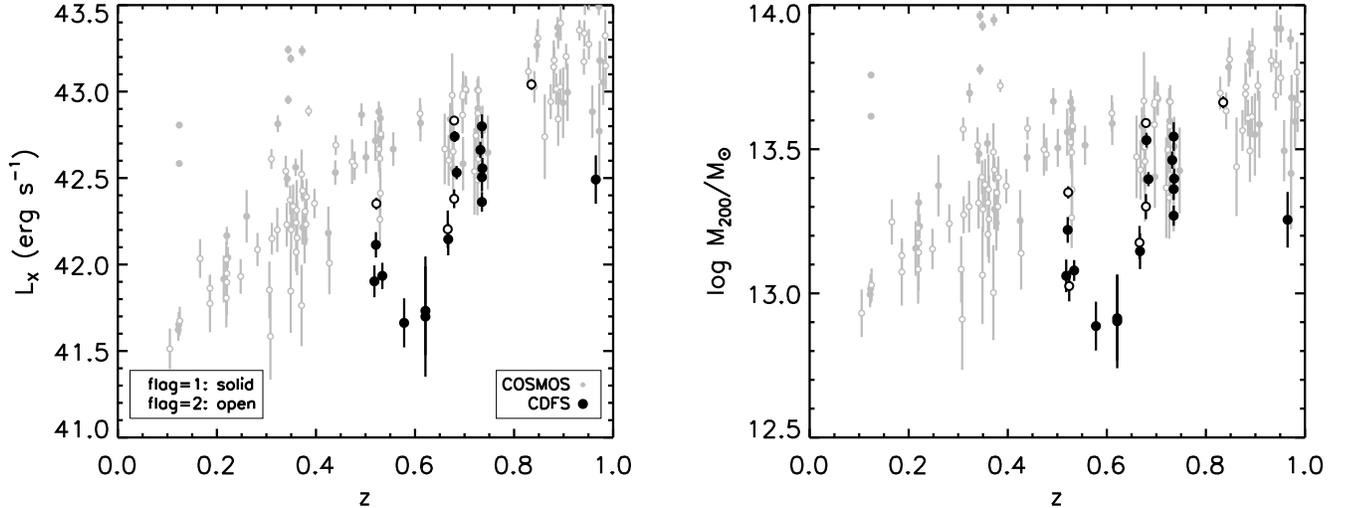}
\caption{Selection properties of the X-ray group sample.  ($a$) Rest-frame X-ray luminosity (0.1-2.4~keV) versus redshift for groups from our primary sample in the deep CDFS field (black) and the comparison sample in the shallower COSMOS field (gray).  Filled circles indicate flag=1 while open symbols indicate flag=2 (see text in Section~\ref{sec_groupsample}).  ($b$) \mh\ versus redshift for the same sample.  The CDFS sample reaches lower halo masses at $z>0.5$, probing below $M<10^{13.3}$~\msun.} \label{fig_selection}
\end{figure*}

For each X-ray group, we assign CSI galaxies above the mass limit within the projected virial radii, $R_{200}$, and within $|\Delta z|/(1+z)<0.02$.  This large redshift window encompasses both members and contaminants (see below for correction).  We sum the stellar masses of the galaxies down to \mgroupmcut, multiplying by the completeness weights.  On spatial scales comparable to those of the groups at \zwindow\ (e.g., $D \sim 1\arcmin$), the completeness correction accurately reproduces galaxy counts to within $\sim 20\%$, an uncertainty that is factored into the total error budget on the aggregate stellar mass measurements.

To account for galaxies below our stellar mass completeness limit, we correct these group stellar masses assuming a Schechter function with $\alpha=-1$ and $\log M^{*}=10.9$.  Integrating down to $M=10^9$~\msun\ we find a correction factor of $1.27$. Changing $\alpha$ and $M^{*}$ within reasonable ranges yields an uncertainty of $\sim 30\%$ in this correction, which is added to the error budget.

We derive a contamination correction for every group by subtracting off the stellar mass contribution from foreground and background galaxies.  We measured the stellar mass within circular apertures of radius $R_{200}$ at spatial locations that did not overlap with other groups near $z_{\rm group}$.  We compute the aggregate stellar mass of galaxies at $|\Delta z|/(1+z_{\rm group})<0.02$ in these apertures and take the mean and standard deviation to represent the background and its uncertainty (typically $\sim 18\%$).

After measuring the stellar mass within $R_{200}$, we correct the value to that expected within $r_{200}$ assuming a spherically symmetric NFW profile \citep{nfw1996} seen in projection \citep[e.g.,][see their Equation~7]{giodini2009}.  We assume an NFW profile with concentration of $c=4$ resulting in a multiplicative correction factor of $\sim 0.78$.  The range of typical concentration values associated with high mass halos \citep[$c \sim 3-5$,][]{duffy2008b} impacts the correction factor by only $\sim 3\%$ and therefore does not significantly affect our results or conclusions.

Lastly, there is an additional source of uncertainty due to the statistical uncertainties in the individual mass measurements.  The typical statistical uncertainty, based on repeat UDP observations and spectrophotometric SED fitting, is $\sim 0.1$~dex per galaxy.  This does not include contributions due to systematic effects such as different stellar population synthesis models and IMF variations.  

The total uncertainty in the aggregate stellar mass measurement for a typical group in our sample is $\sim 0.2$~dex.

Figure~\ref{fig_map} shows two example low-mass groups in the CDFS.  Color images were constructed using HST CANDELS WFC3 $J_{125}$ and $H_{160}$ imaging \citep{grogin2011,koekemoer2011}.  Within the virial radii (dashed green circles), Figure~\ref{fig_map} shows clear overdensities of group members (blue circles) relative to field galaxies within the same redshift slice (red and orange circles), indicating that we are able to reliably place galaxies into groups with our data and compute robust aggregate stellar mass totals.

\subsection{COSMOS Comparison Sample} \label{sec_cosmos}

We compare our results for low-mass groups at $0.5<z<1$ to low-mass groups at $0.1<z<0.5$ and to high-mass groups at $0.5<z<1$, both derived from the wider field, but shallower X-ray data from COSMOS \citep[for a detailed study, see][]{giodini2009}.  We standardize the group stellar masses in COSMOS with ours in the CDFS by measuring them in the same manner as described in the previous section and using the photometric redshifts and stellar masses from \citet{george2011}.  While this approach differs from summing redshift likelihood functions within the redshift windows of the groups, as carried out in \citet{george2011}, there is very good agreement between the two methods with a scatter in the aggregate stellar masses of only $\sim 0.06$~dex and a negligible offset.  We convert the COSMOS stellar masses from Chabrier to Kroupa IMF by adding $0.05$~dex.  In total, the COSMOS sample consists of $125$ groups, $69$ at $0.5<z<1$ and $56$ at $0.1<z<0.5$ with halo masses spanning $10^{12.8}<M_{200}/M_{\odot}<10^{14.3}$ and with X-ray flag=1 or flag=2 and no other flags.

\section{Results}

\begin{figure*}
\epsscale{1.15}
\plottwo{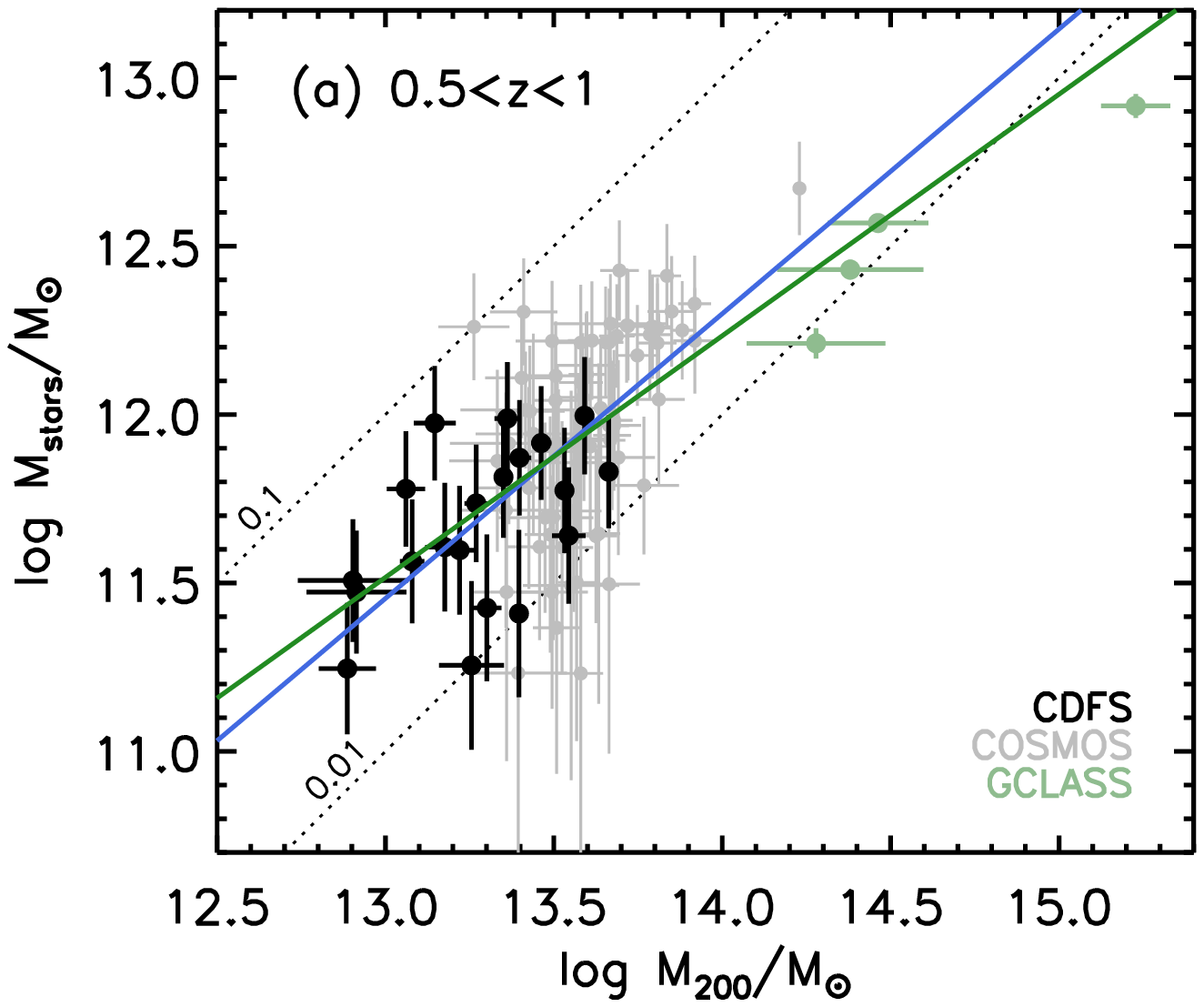}{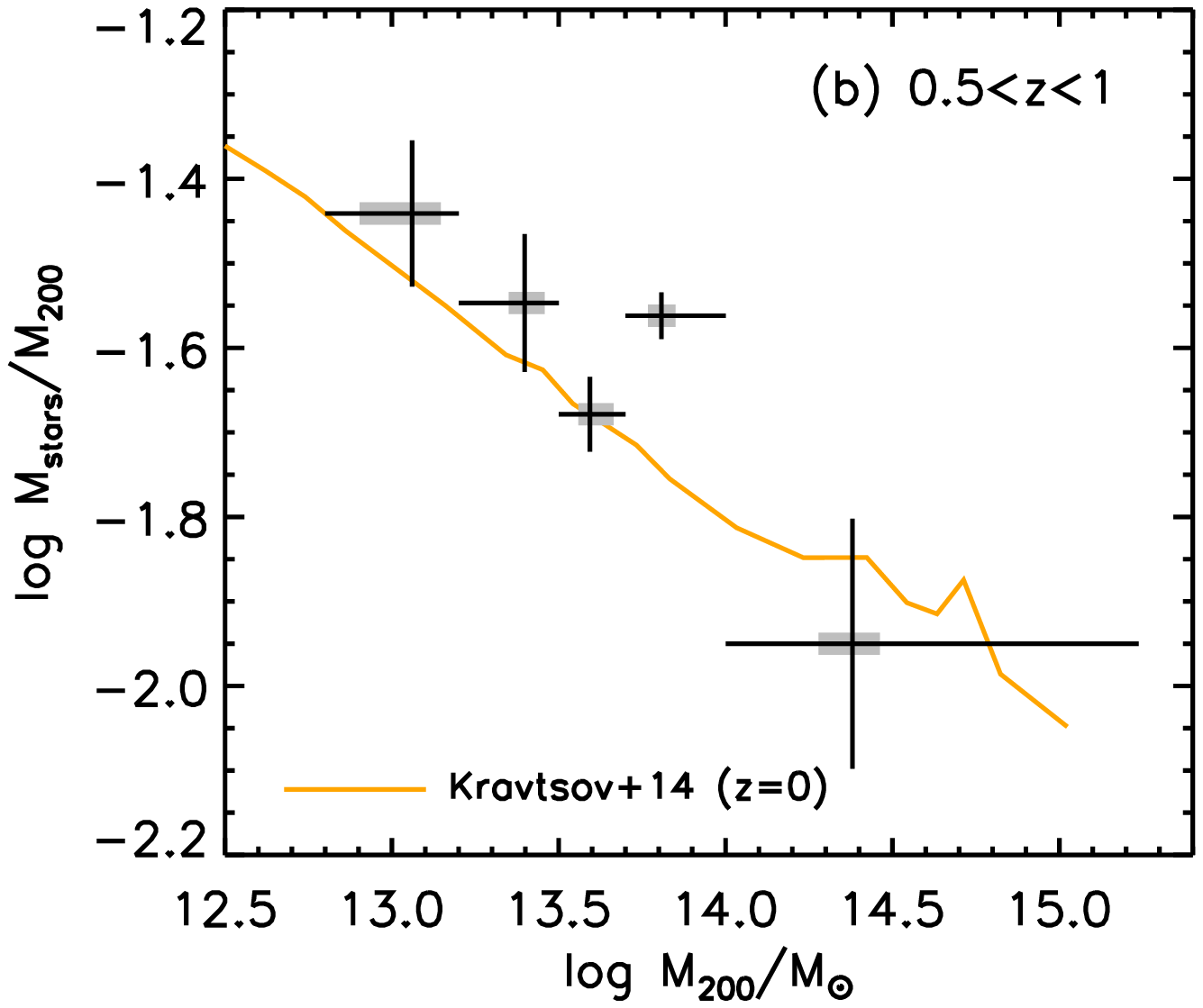}
\caption{The stellar mass-halo mass relation (\mgroup\ versus \mh) for high mass halos at $0.5<z<1$.  Stellar mass measurements represent the aggregate mass in stars of all galaxies in the halo. ($a$) Individual X-ray detected groups in the CDFS and COSMOS are in black and gray, respectively.  Clusters from GCLASS are in green \citep{vanderburg2014}.  Measurements and uncertainties from the three datasets have been standardized when possible.  Dotted lines across the figure indicate ratios of stellar-to-halo mass of $0.01$ and $0.1$ for reference.  A biweight fit to the CDFS+COSMOS data (blue line) indicates a slope of $0.84 \pm 0.10$ and an observed scatter of $\sigma=0.25$~dex.  Fitting all of the data, including GCLASS (green line), the slope becomes $0.72 \pm 0.07$ and the observed scatter $\sigma = 0.25$~dex.  The less-than-unity slope indicates that fewer stars are formed/assembled in higher mass halos.  The relatively small observed scatter implies that total stellar mass serves as a good proxy for the total halo mass in this regime.  ($b$) Median stellar to halo mass ratio for different bins in halo mass.  Vertical error bars represent bootstrapped uncertainties.  Black horizontal error bars indicate the bin range while the thicker gray error bars represent the 25-75th percentile values for the halo masses in each subsample.  At low group masses of $10^{12.8}<M/M_{\odot}<10^{13.5}$ probed mostly by the data in the CDFS, stars make up $3-4\%$ of the total mass of the halos.  The orange curve indicates the $z \sim 0$ relation from abundance matching.
} \label{fig_stellar_halo_mass}
\end{figure*}

\subsection{Galaxy Group Sample Properties} \label{sec_groupsample}

Figure~\ref{fig_selection}$a$ shows the X-ray luminosity of X-ray groups as a function of redshift.  Redshifts were determined from followup spectroscopy of red sequence candidate galaxies as described in \citet{george2011} and \citet[][submitted]{finoguenov2014}.  The filled symbols represent X-ray groups with centroids determined from the X-ray image (flag=1), while the open symbols represent centroids determined from the positions of the red sequence galaxies (flag=2).  The median redshift for the CDFS sample is $z_{\rm med} \sim 0.7$.  Figure~\ref{fig_selection}$b$ shows the COSMOS and CDFS group masses, \mh, as a function of redshift.  The error bars represent the statistical uncertainty and do not include the contribution due to the scatter in the $L_X-M$ relation ($\sim 0.2$~dex).  As a result of the lower flux limits, CDFS reaches significantly lower group masses at high redshift compared to COSMOS.  In particular, halos at $z>0.5$ with mass below $M_{200}<10^{13.3}$~\msun\ are almost exclusively detected in the CDFS.

\subsection{The Stellar Mass -- Halo Mass Relation at \zwindow}

The aggregate stellar mass within X-ray groups at \zwindow\ is compared to their total mass in Figure~\ref{fig_stellar_halo_mass}.  In Figure~\ref{fig_stellar_halo_mass}$a$, the CDFS groups are shown using the black points. The COSMOS groups are shown in light gray.  Owing to its depth, the CDFS sample provides the only leverage on $M_{\rm stars}/M_{200}$ for low mass groups ($M_{200}<2\times10^{13}$~\msun).

The blue line is a biweight fit to the CDFS and COSMOS data of the form

\begin{eqnarray}
\log M_{\rm stars}= \beta + \alpha \log M_{200} \label{eq_cc}
\end{eqnarray}
where $\alpha=0.84 \pm 0.10$ and $\beta=0.47 \pm 1.33$ (see Table~\ref{table_fits}).  The slope is consistent with the value of $\alpha=0.81 \pm 0.11$ found in \citet{giodini2009} for their higher mass COSMOS group sample spanning $0.1<z<1$.  The implied stellar fraction for the COSMOS groups ($\sim 2\%$ at $M_{200} \sim 10^{13.5}$~\msun) is also consistent with the results of \citet{leauthaud2012b}.  The green data points in Figure~\ref{fig_stellar_halo_mass} represent measurements for clusters from the GCLASS survey \citep{vanderburg2014}.  The green line is a fit to all three datasets and results in $\alpha=0.72 \pm 0.07$ and $\beta=2.20 \pm 0.97$.  The inclusion of the cluster data results in a shallow slope which is consistent with the general trend found at high halo masses.

The observed biweight scatter about the relation in Equation~\ref{eq_cc} is $\sigma=0.25$~dex, and is comparable for both the CDFS and COSMOS samples.  The similar scatter is reassuring and implies that our completeness corrections are robust and unbiased.  Furthermore, this observational scatter is remarkably small.  When factoring in the uncertainties on the aggregate stellar mass measurements, the intrinsic scatter must be quite low.  Determination of the intrinsic scatter requires accurate estimates of the observational errors. Because they total $\sim 0.2$~dex, we are unable to make a precise determination of the intrinsic scatter.  In any case, an upper limit for the intrinsic scatter in the group stellar mass versus halo mass relation of $\sim 0.25$~dex reflects a relatively tight correlation suggesting that the former serves as a rough proxy for the latter.  This has also been noted by other authors \citep[e.g.,][]{vanderburg2014}.

Figure~\ref{fig_stellar_halo_mass}$b$ shows the median stellar to halo mass for different bins in halo mass.  The bulk of the low mass CDFS group sample lies in the two lowest mass bins spanning $10^{12.8}<M_{200}/M_{\odot}<10^{13.5}$. In these halos, which are 10-20 times more massive than the Milky Way's halo, the stellar mass constitutes $3-4\%$ of the total mass budget.

\subsection{No Significant Evolution in the Group Stellar to Halo Mass Ratio since $z \sim 1$}

\begin{figure}
\epsscale{1.25}
\plotone{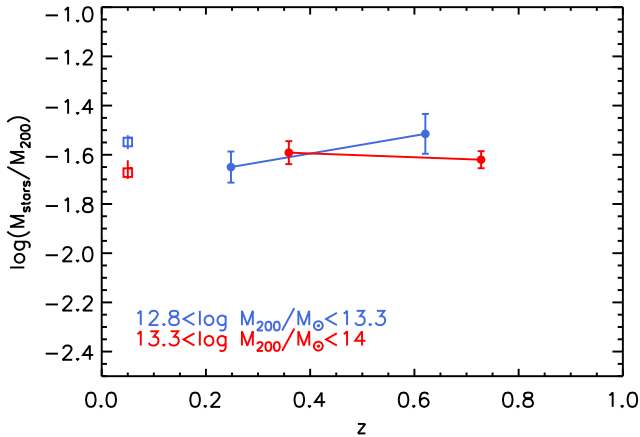}
\caption{Ratio of group stellar mass to halo mass versus redshift for two bins of halo mass.  Low mass halos ($10^{12.8}<M/M_{\odot}<10^{13.3}$) are shown in blue and high mass in red ($10^{13.3}<M/M_{\odot}<10^{14}$).  Results from abundance matching at $z \sim 0$ \citep{kravtsov2014} are shown as open squares.  We do not find any significant evolution in the ratio of stellar to halo mass for groups at $z \lesssim 1$.} \label{fig_evolution}
\end{figure}

\begin{deluxetable}{lcccc}
\tablewidth{0pc}
\tablecolumns{5}
\tablecaption{Best fit parameters to stellar-halo mass relation}
\tablehead{
  \colhead{Sample Fitted} & \colhead{$\alpha$} & \colhead{$\beta$} & \colhead{$\sigma_{obs}$} & \colhead{$N$} \\ 
  \colhead{} &  \colhead{} &  \colhead{} &  \colhead{(dex)} &  \colhead{} \label{table_fits}
} 
\startdata
CDFS, COSMOS & $0.84 \pm 0.10$ & $0.47 \pm 1.33$ & $0.25$ & 89 \\
CDFS, COSMOS, GCLASS & $0.72 \pm 0.07$ & $2.20 \pm 0.97$ & $0.25$ & 93 
\enddata
\tablecomments{The fit coefficients $\alpha$ and $\beta$ are indicated in Equation~\ref{eq_cc}: $\log M_{\rm stars}=\beta + \alpha \log M_{200}$.  The observed biweight scatter about this relation is $\sigma_{obs}$ and the number of halos included in the fit, $N$.  Fits apply only to the mass range probed by the data presented in this work.}
\end{deluxetable}

Here we examine the redshift evolution, if any, of the stellar mass -- halo mass relation for groups.  The orange curve in Figure~\ref{fig_stellar_halo_mass}$b$ indicates the stellar-to-halo mass ratio from abundance matching, empirically derived by \citet[][Figure~12]{kravtsov2014} at $z \sim 0$.  Given the error bars and any additional uncertainty due to systematic differences between the two measurement methods, there does not appear to be any significant evolution at $z \lesssim 1$ in the stellar-to-halo mass relation for $M_{200} \gtrsim 10^{13}$~\msun.

We can further examine and confirm the lack of evolution by using additional measurements at lower redshifts ($0<z<0.5$) that were made in the same manner as our own.  Figure~\ref{fig_evolution} shows the redshift evolution for the median stellar-to-halo mass ratio for two halo mass bins.  Uncertainties were computed by bootstrapping.  Note that in the lowest mass bin, $10^{12.8}<M_{200}/M_{\odot}<10^{13.3}$ (blue), the entire sample at $z>0.5$ is comprised of groups in the CDFS where the X-ray imaging is of sufficient depth to detect these systems and stellar masses have been measured with CSI.  The low redshift ($z<0.5$) sample is comprised exclusively of COSMOS X-ray groups (see Section~\ref{sec_cosmos}).  The open squares indicate the results at $z \sim 0$ from abundance matching \citep{kravtsov2014} and the associated error bars the range of median values spanned by halos with masses in the 25-75th percentile of the two sub-samples.  Again, we do not find any significant evolution in the stellar-to-halo mass ratio in the group regime at $z \lesssim 1$.

\section{Discussion}

We have measured the stellar mass -- halo mass relation for low mass groups at $0.5<z<1$, complementing other work at these redshifts \citep[e.g.,][]{giodini2009,vanderburg2014}.  For the low mass groups in our CDFS sample, we find that about $\sim 3-4\%$ of the total mass is contained in stars, less than that of $L^{\star}$ haloes but more than that of clusters, as is also the trend in the local universe.  The lack of (or slow) evolution in the stellar fraction of such halos is relevant given that the number density of these groups is observed to increase dramatically at $z<1$ \citep{williams2012b} indicating the growing significance of such environments.  

Late-time formation of massive halos is expected from hierarchical growth.  Massive halos in the mass range $M_{200}>10^{13}$~\msun\ are predicted to comprise $\sim 32\%$ of the total mass of all halos above $M_{200}>10^{11.6}$~\msun\ (approximately the halo mass scale that would host central galaxies at our stellar mass limit) at $z \sim 1$ \citep[based on HMFcalc,][]{murray2013,tinker2008c}.  By $z \sim 0$, this percentage rises substantially to $\sim 52\%$.  For groups in the mass range $10^{13}<M_{200}/M_{\odot}<10^{14}$, the percentage increases from $\sim 29\%$ at $z \sim 1$ to $\sim 36\%$ at $z \sim 0$.  Clusters at $M_{200}>10^{14}$~\msun\ clearly contribute toward the mass growth at $z<1$ but even these structures are built, in part, from groups \citep{berrier2009}.

The increase in number densities of galaxy groups, and the associated increase in the stellar mass density within such halos, has implications for the buildup of quiescent galaxies (QGs) since $z \sim 1$.  From $z \sim 1$ to $z \sim 0$ the quiescent fraction (QF) increases from $\sim 44\%$ to $\sim 59\%$ for galaxies with \masslimit\ \citep{kelson2014,muzzin2013c}.  Down to this mass limit, QFs are higher in rich environments compared to the field \citep{dressler1980} with QFs of $\sim 80\%$ and $\sim 35\%$ in groups/clusters and the field, respectively \citep[e.g.,][]{patel2011}.  As a consequence, the rise of groups over the past $\sim 8$~Gyr contributes substantially to the evolving QF over that time.  By assessing the star formation activity and quiescence of galaxies in groups over the full 15~degs$^2$ survey field of CSI, we will accurately constrain the impact of large scale structure growth on the buildup of QGs and the decline in global SFRD since $z\sim 1$.

\acknowledgements
We thank Alexis Finoguenov for making his CDFS X-ray catalog available.



\end{document}